\documentstyle[11pt,baaa-eng,twoside,epsf,psfig]{article}
\begin{document}
\vskip 1.0cm
\markboth{M.E. De Rossi, P.B. Tissera \& C. Scannapieco}{Luminosity-Metallicity-Stellar Mass correlation in galaxies}
\pagestyle{myheadings}


\vspace*{0.5cm}
\parindent 0pt{ COMUNICACI\'ON DE TRABAJO -- CONTRIBUTED  PAPER } 
\vskip 0.3cm
\title{Study of the Origin of the Luminosity-Metallicity and the Stellar Mass-Metallicity Relations in Hierarchical Universes} 

\author{Mar\'{\i}a Emilia De Rossi}
\affil{Instituto de Astronom\'{\i}a y F\'{\i}sica del Espacio, Ciudad de Buenos Aires, Argentina, derossi@iafe.uba.ar}

\author{Patricia Beatriz Tissera}
\affil{Instituto de Astronom\'{\i}a y F\'{\i}sica del Espacio, Ciudad de Buenos Aires, Argentina, patricia@iafe.uba.ar}

\author{Cecilia Scannapieco}
\affil{Instituto de Astronom\'{\i}a y F\'{\i}sica del Espacio, Ciudad de Buenos Aires, Argentina, cecilia@iafe.uba.ar}

\begin{abstract} In this work, we study the Luminosity-Metallicity relation (LMR) and the Stellar Mass-Metallicity relation (MMR) of
galactic systems in a hierarhical clustering scenario. We performed  numerical hydrodynamical simulations with
the chemical {\small GADGET-2} of Scannapieco et al.(2005) in a $\Lambda$CDM universe.
 We found that our simulated galactic systems reproduce the observed local LMR and its evolution in zero point and slope.
The simulated MMR is also in agreement with recent 
observational results. From the analysis of the evolution of the MMR, we found a 
characteristic mass at $\approx 10^{10.2} M_{\odot} \, h^{-1}$
  which   separates two galactic populations with different astrophysical properties. More massive systems tend to have their stars formed at $z > 2$
and show less evolution than smaller systems. Hence, this characteristic mass is determined by the formation of the structure in
a hierarchical scenario. Our results also suggest the need for efficient supernova feedback. 
\end{abstract}

\begin{resumen}  
En este trabajo, estudiamos la relaci\'on Luminosidad-Metali\-cidad (LMR) y 
la relaci\'on Masa Estelar-Metalicidad (MMR) de los sistemas gal\'acticos 
en un modelo de agregaci\'on jer\'arquica.  Realizamos simulaciones
num\'ericas hidrodin\'amicas con el c\'odigo qu\'{\i}mico {\small GADGET-2} 
de Scannapieco et al. (2005) en un universo $\Lambda$CDM.  Encontramos
que nuestros sistemas gal\'acticos simulados reproducen la LMR local y 
su evoluci\'on en el punto cero y la pendiente.  La MMR simulada
est\'a tambi\'en en acuerdo con resultados observacionales recientes.  
A partir del an\'alisis de la evoluci\'on de la MMR, hallamos una
masa caracter\'{\i}stica en $\approx 10^{10.2} M_{\odot} \, h^{-1}$, la cual
separa dos poblaciones gal\'acticas con diferentes propiedades 
astrof\'{\i}sicas.  Los sistemas m\'as masivos
tienden a formar sus estrellas a $z>2$ y muestran menor evoluci\'on que los sistemas
peque\~nos.  Entonces, esta masa caracter\'{\i}stica es determinada
por la formaci\'on de la estructura en un universo jer\'arquico.
Nuestros resultados sugieren tambi\'en la necesidad de un importante
{\it feedback} por supernovas.
\end{resumen}

\section{Introduction}

Determining the chemical composition of galaxies is of fundamental 
importance for tracing back the history of evolution
of galaxies. 
In particular, the LMR has been widely studied in the Local 
Universe.  Metallicities are tightly related 
with the luminosities of galaxies in such a way, that brighter systems have higher abundances (Lamareille et al. 2004). 
 Furthermore, recent studies have also suggested that this relation extends to intermediate redshifts but displaced towards lower metallicities and higher luminosities (Kobulnicky et al. 2003).

When studying galaxy evolution, stellar mass is a better 
parameter than luminosity.  However, because of the 
difficulties in obtaining stellar masses, most studies 
have used luminosity as a surrogate.  Recently, though, 
Tremonti et al. (2004) have estimated the relation 
between metallicity and stellar mass in the Local Universe.  
The authors found a strong correlation extended over 2 dex in stellar mass and a factor of 10 in metallicity.

In this work, we study the evolution of the MMR and the LMR by 
employing numerical chemo-dynamical simulations which allow to describe the non-lineal growth of structure simultaneously with the enrichment of the interstellar medium in a cosmological framework.

\section {Results and discussion}

We have run numerical simulations by using the chemical 
{\small GADGET-2} of Scannapieco et al. (2005).  
A $\Lambda$CDM cosmological model ($\Omega$=0.3, $\Lambda$=0.7, 
$\Omega_{b}$=0.04 and $H_{0}$=100
 $h$ km s$^{-1}$ Mpc$^{-1}$ with $h$=0.7) was assumed, 
according to which galaxies formed by the hierarchical 
aggregation of substructures.
We have analysed two realizations of the
power spectrum in a 10 Mpc $h^{-1}$ side box,
 initially resolved with 
$2 \times 160^3$ (S160) and $2 \times 80^3$ (S80) particles, 
corresponding to mass resolutions of 
$2.71 \times 10^6$ M$_{\sun} \, h^{-1}$ and 
$2.17 \times 10^7$ M$_{\sun} \, h^{-1}$ 
for the gas phase and 
$1.76 \times 10^7$ M$_{\sun} \, h^{-1}$ and 
$1.41 \times 10^8$ M$_{\sun} \, h^{-1}$ 
for dark matter respectively.

A Salpeter Initial Mass Function has been assumed 
with upper and lower limits of 40 M$_{\sun}$ and 
0.1 M$_{\sun}$, respectively. 
 The chemical model includes metal-dependent radiative 
cooling, star formation and chemical enrichment by supernovae II and Ia (Scannapieco et al. 2005).

Galactic objects were identified by applying an identification
algorithm that combines the friends-of-friends 
technique and the contrast density criterium of 
White, Efstathiou \& Frenk (1993). 
Dynamical and chemical properties were estimated at
 the optical radius 
calculated accordingly to the standard definition
 as the radius  which contains the 83\% of the 
baryonic mass of the system
(Tissera et al. 2005).  Colours and magnitudes of 
galactic systems were calculated by resorting to  
population synthesis models (see De Rossi et al. 2006
in preparation).

Our simulations predict a linear correlation 
between oxygen abundance and luminosity 
which is in good  agreement with the observational results.  
We have also found an evolution in the LMR in such a way that 
the slope increases and the zero point decreases with redshift
consistently with the findings of
Kobulnicky \& Kewley (2004), among others.  In particular, we decided to 
work with the I-band because it is less affected by
extinction and can be more directly related
with the underlying mass distributions.  Our
results indicate  that
at a given chemical abundance, galactic systems are 
$\sim$ 3 dex brighter at
$z=3$ compared to $z=0$, with the larger
 evolution at fainter
magnitudes. Futhermore, we have encountered a mean evolution
in the chemical abundances of galactic
systems of $\sim$ 1.6 dex for
brighter magnitudes and  $\sim$ 2.5 dex for faint ones,
from $z=3$ to $z=0$.

We have also analysed the MMR for simulated galactic systems, 
obtaining similar trends to those found by 
Tremonti et al. (2004) in the Sloan Digital Sky Survey (SDSS)
 but with a displacement of -0.25 dex in the zero point.
  This last difference  may be explained taking into account 
that the SDSS explored only the central regions of galaxies 
which could lead 
 to an overestimation of their metal content.
Galactic abundances derived from simulations tend to 
increase with stellar mass which is also
consistent with the observed behaviour. However, we obtained an 
excess of metals in the lower mass end which could be 
due to the absence of supernovae energy feedback in our model. 

We have determined a characteristic stellar mass at
$M_{c}=10^{10.2}$ M$_{\sun} \, h^{-1}$ where a change 
in the curvature of the MMR occurs. This characteristic mass,
 which corresponds to an oxygen abundance of $\sim$ 8.7 dex, 
 has been obtained by estimating where 
the residuals of the linear fits depart systematically from zero. It is important to note that this mass is similar to the characteristic mass derived from the SDSS by Tremonti et al. (2004) and Gallazzi et al.(2005).

In addition, we have found that the MMR 
 exhibits the same general patterns from $z=3$ to $z=0$, 
but with a displacement towards higher abundances as redshift decreases.
The          
characteristic stellar mass $M_c$ remains almost unchanged with time 
and only its corresponding chemical abundance evolves by
0.05 dex in the same redshift range.
The major departure from the local MMR occurs for smaller
systems which
increase their chemical content by $\sim$ 0.10 dex.  On the 
other hand, massive systems show less
evolution with variations of $\sim$ 0.05 from
$z=3$ to $z=0$.

We have also studied the Metallicity-optical Velocity Relation (MVR)
finding a well defined correlation from $z=3$ to $z=0$.
However, a higher level of evolution has been found in the MVR when 
compared to the MMR.  Fast rotators  show an
enrichment of  $\sim$ 0.18 dex from $z=3$ to $z=0$ while at lower
metallicities the variations are of $\sim$ 0.28 dex. 
This significant evolution of the MVR is a 
consequence of the increase of the mean density of the Universe 
at high redshift, so that at a fix stellar mass, systems need 
to concentrate more  in order to fullfill the contrast 
density criterium and, hence, galaxies reach higher velocities.  
The larger evolution showed by the simulated MVR, when compared with the MMR, shows
 that the second is more fundamental in the sense that it is not strongly dependent
on the cosmic epoch (see Tissera, De Rossi \& Scannapieco 2005 for
details).

By analysing the merger trees of systems at $z=0$, 
we have encountered that the features of the MMR can be 
traced back in time taking into account the mergers and interaction
history 
of galaxies within the hierarchical aggregation picture.
 More massive systems transform most of their gas into 
stars at high redshifts and experience important mergers.
 At lower redshifts, these galactic objects are saturated 
of stars, so that, while their stellar mass largely increases
in a merger event, their metallicity remains basically 
unchanged.
On the other hand, less massive systems form their stars in a more passive way or by rich-gas mergers leading to a more tight correlation between metallicity and stellar mass.

\section{Conclusions}

We have estimated the MMR, the LMR and the MVR correlations
 finding
similars trends to those detected in observations.
These relations evolve with redshift with
the major changes driven by less massive systems
in consistency with the downsizing scenario.

We have found an excess of metals for
lower masses which could be attributed
to the lack of an energy feedback treatment in our
simulations.

A characteristic stellar mass $M_c \sim 10^{10.2} M_{\sun} h^{-1}$ has been determined
at which the MMR flattens.  This mass is in 
good agreement with the one found by
Tremonti et al (2004) and Gallazzi et al. (2005).

Our results suggest that the features of the MMR
are tightly related to the hierarchical aggregation
scenario in which galaxies form.

\acknowledgments 
We are grateful to CONICET, Fundaci\'on Antorchas and LENAC.
The simulations were performed on Ingeld PC cluster and
on HOPE cluster at IAFE (Argentina).


\begin{references}
\reference  Gallazzi, A., Charlot, S., Brinchmann, J., White, S. D. M., Tremonti, C., 2005, MNRAS, accepted (astroph/0506539)
\reference  Kobulnicky, H. A., Willmer, C. N. A., Phillips, A. C., Koo, D. C., Faber, S. M. et al. 2003, ApJ 599, 1006
\reference  Kobulnicki, H. A., Kewley, L. J. 2004, ApJ  617, 240
\reference  Lamareille, F., Mouhcine, M., Contini, T., Lewis, I., Maddox, S. 2004, MNRAS, 350, 396
\reference  Scannapieco, C., Tissera, P. B., White, S. D. M., Springel, V., 2005, MNRAS, accepted (astroph/0505440)
\reference Tissera, P. B., De Rossi, M. E., Scannapieco, C., 2005, MNRAS, 364, 38L
\reference  Tremonti, C. A., Heckman, T. M., Kauffmann, G., Brinchmann, J., Charlot, S. et al. 2004, ApJ 613,898
\reference  White, S. D. M., Efstathiou, G. \& Frenk, C. S., 1993, MNRAS, 262, 1023
\end{references}
\end{document}